\documentclass[]{myaa}
\usepackage{natbib}
\usepackage{txfonts}
\usepackage{balance}
\usepackage{graphicx}

\usepackage[letterpaper]{hyperref}

\bibpunct{(}{)}{;}{a}{}{,} 
\begin{document}
\def\teff{$T\rm_{eff }$}
\def\kms{${\mathrm {km s^{-1}}}$}
\idline{1}{1}

\title{Variations in the  lithium abundances of turn off
stars in   the globular cluster 47 Tuc 
\thanks{Based on
observations made with the 
ESO VLT-Kueyen telescope at the Paranal Observatory, Chile,
in the course
of the ESO-Large program 165.L-0263
                            }}

   \subtitle{}

\author{
P. \,Bonifacio \inst{1,2,3}
\and
L. Pasquini\inst{4}
\and
P. \,Molaro  \inst{2,3}
\and
E. Carretta\inst{5}
\and
P.~Fran\c cois\inst{2}
\and
R.G. Gratton\inst{6}
\and
\hbox{G. James\inst{4}}
\and
L. Sbordone\inst{1,2,3}
\and
F.~Spite\inst{2}
\and
M. Zoccali\inst{7}
          }
\offprints{P. Bonifacio}

\institute{
CIFIST Marie Curie Excellence Team
\and
GEPI, Observatoire de Paris, CNRS, Universit\'e 
Paris Diderot ; Place Jules Janssen 92190
Meudon, France
\and
Istituto Nazionale di Astrofisica, 
Osservatorio Astronomico di Trieste,  Via Tiepolo 11, 
I-34131 Trieste, Italy 
\and
European Southern Observatory, Karl--Schwarzschild--Str. 2, D-85748 
Garching bei M\"unchen, Germany
\and
Istituto Nazionale di Astrofisica, 
Osservatorio Astronomico di Bologna, via Ranzani 1, 40127
Bologna, Italy
\and
Istituto Nazionale di Astrofisica, 
Osservatorio Astronomico di Padova, Vicolo dell'Osservatorio 5, I-35122
 Padova, Italy
\and
Department of Astronomy, Pontificia Universidad Catolica,
Av. Vicu\~na Mackenna 4860, Santiago 22, Chile
}

\authorrunning{Bonifacio et al.}
\mail{P. Bonifacio}

\titlerunning{Lithium in 47 Tuc}

\date{Received  / Accepted }

\abstract
{}
{ Our aim is to determine  Li abundances in TO stars of
the Globular Cluster 47 Tuc and test theories
about Li variations among TO stars. }
{
We make use of high resolution (R$\sim 43000$), 
high signal-to-noise ratio (S/N=50--70)
spectra of 4 turn off (TO) stars 
obtained with the UVES spectrograph at the
8.2m VLT Kueyen telescope.}
{
The four stars observed,  span the range
$1.6\la {\rm A(Li)} \la 2.14$, providing 
a mean A(Li) = 1.84 with a standard deviation
of 0.25 dex. When coupled with data of other
two TO stars of the cluster,  available in the literature,
the full range in Li abundances 
observed in this cluster is $1.6\la {\rm A(Li)} \la 2.3$.
The  variation in A(Li) is at least 
0.6 dex (0.7 dex considering also
the data available in the literature)
and the scatter is  six times 
larger than what expected from the
observational error. We 
claim that  these variations are real.
A(Li)  seems to be anti-correlated with A(Na) exactly
as observed in  NGC 6752. 
No systematic error in our analysis could produce such
an anti-correlation. 
}
{
Na production through $p$ captures on $^{22}$Ne
at temperatures in excess of $3\times 10^7$K and the 
contemporary  Li destruction could result  in this
anti-correlation. However such nuclear processing 
cannot have taken place in the stars themselves,
which do not reach such high temperatures, even at their
centre. This points towards the processing 
in a previous generation of stars.
The low N/O ratios in the observed 
stars
and the apparent lack of correlation
between N an Li abundances, place a strong
constraint on the properties of this previous
generation.
 Our results indicate a different behaviour among the
Globular Clusters so far studied  as
far as the abundance patterns are concerned.
}
\maketitle           

\section{Introduction}

The Globular Cluster (GC)
47 Tuc is a good example of the metal-rich end population 
of these very old objects (age in the range
 11 to 14  Gyr,
\citealt{gra03}). 
In view of its brightness it is one of the best studied
GCs and with the advent of the UVES spectrograph
at the ESO-VLT it has become possible to
obtain abundances of individual stars  on the Main Sequence,
with an accuracy previously possible only 
for Halo field stars which are several 
magnitudes brighter.
It  is one of the main targets of the ESO Large Programme
165.L-0263 (P.I. R.G. Gratton). The chemical composition
of turn off (TO) and subgiant (SG)  stars 
from our  UVES data is given in another paper
of this series \citep{Carretta}, which provided a
metallicity [Fe/H]=--0.64 for the TO stars of this cluster.

It has been known for almost thirty years
that 47 Tuc exhibits a bimodal distribution
of CN band strengths, suggesting a
bimodal distribution of N abundances, among
giant stars \citep{norris79} and also among
TO stars \citep{briley}. Thus abundance inhomogeneities
among stars in this cluster are expected.

In this paper we examine the Li abundances in the TO stars.
The only previous investigation of Li in this cluster
was performed by \citet{pm97} who could detect the
Li line in two TO stars, using spectra
obtained with EMMI on the 3.5m ESO NTT telescope. 
In the present  investigation we shall make
use of the equivalent widths measured by \citet{pm97}.

Old metal--poor warm dwarfs in the Galactic Halo
show a rather uniform lithium abundance,
whichever their metallicity or effective temperature.
This  was discovered by  \citet{spite82}
and is usually called the {\em Spite Plateau}.
This behaviour of lithium is unique among
chemical  elements, all of which show
a decreasing abundance with decreasing 
metallicity; lithium is the only element to display
a plateau. 
The most obvious interpretation is still
the one put forward by \citet{spite82}
that the lithium observed in the {\em Spite Plateau}
is simply the ``primordial'' lithium, that
is the lithium that has been produced during
the big bang, together with the other
light nuclei, D, $^3$He and $^4$He.
The abundance of the nuclei produced
in this way depends on the baryon to
photon ratio ($\eta = n_b/n_\gamma$)
which cannot be deduced from first principles,
but has to be somehow measured.
The WMAP satellite  has provided 
this ratio with an accuracy of the order
of 4\% \citep{spergel,spergel06},
the precise value being
$\eta = 6.11\pm 0.22\times 10^{-10}$.
When inserted in standard big bang
nucleosynthesis computations
(SBBN) this value implies
A(Li)\footnote{Throughout the paper we use the notation
A(X) = log[N(X)/N(H)] +12.} = 2.64.
Current estimates of the level of the
{\em Spite Plateau} range from 2.1
\citep{B06} to 2.3 \citep{B02,melendez}.
There is thus tension between the
observed lithium abundances in stars
and the predictions of SBBN, when the 
baryonic density derived from WMAP is adopted.
The most obvious
ways to reconcile these results are either
to look for new physics at the time of
nucleosynthesis or to find mechanism(s)
which have depleted uniformly Li from
the primordial value to what is currently observed in 
Halo stars.

In order to test the theories which predict Li depletion,
GCs are, in principle,  
an ideal target: the stars have the same age and chemical
composition, at variance with what happens with field stars.
Effects of Li depletion could be obscured by other
concurrent metallicity or age effects.
The Li abundance in GCs is therefore of great importance; 
we may try to detect
some of the features which are predicted by models, such as a mild
scatter in Li abundances, above what is expected from observational
errors, or the existence of ``outliers'', i.e. heavily depleted stars.

Our observations of the GC NGC 6397 \citep{B02} show that
all the observed stars in this metal poor cluster
share the same abundance and there are
no ``outliers''. There is very little
room for intrinsic scatter; this does not rule out any
of the depletion models, but does place a very strong constraint
to be fulfilled. Therefore the currently available observations
seem to argue {\em against} any Li depletion in NGC 6397. 
Our result has recently been challenged
by \citet{Korn06} who have claimed to have found 
a difference in lithium content between 
the TO stars and subgiant stars at an effective
temperature around 5800 K.
According to their interpretation these observations
are in agreement with the diffusive 
models of \citet{richard}.
It should however be kept in mind that the result
of \citet{Korn06} depends on their adopted temperature
scale and that an increase of only 100 K of the
temperatures of the TO stars, as suggested by the
cluster photometry, would erase this difference.
Note that such an increase in the \teff ~ 
would at the same time erase the claimed ``diffusion
signatures'' also in Fe, Ca and Ti. 
The question is therefore still not settled.

At variance, the GC NGC 6752, which is about  a factor
of 4 more metal rich than NGC 6397, displays a strong
variation (up to 0.4 dex )
in Li abundances among TO stars \citep{Pasquini6752}.
These variations, however, do not appear to be random,
but are anti-correlated with the variations of sodium and nitrogen,
and correlated with the variations of oxygen, in the same stars.
Such variations cannot  
be produced by diffusion mechanisms, since the effect of diffusion
would be similar for lithium and sodium.
Neither can they
arise from mixing 
occurring in the stars themselves, since the base of the
convection zone in such stars attains a temperature
of 1.5 MK, which is very far from the region of lithium 
burning (Piau 2005, private communication).
\citet{Pasquini6752} suggest that the most likely source
of such anomalies are intermediate mass asymptotic
giant branch (IM-AGB) stars which have polluted
either the material out of which the TO stars were formed,
or their atmospheres. The nucleosynthetic signatures
of IM-AGB stars are in qualitative agreement with the
observed patterns, although  quantitatively, none of the current
models is capable of fully explaining the observations.
We refer the reader to the papers of
\citet{fenner,ventura05,dantona} and \citet{ventura06}
where some of the problems of IM-AGB models
in reproducing the observed abundance variations
in GCs are discussed, as well as
possible solutions.
From a different perspective \citet{prantzos}
examined the constraints on the cluster's Initial
Mass Function in the case the polluters are IM-AGB
stars or the winds of massive stars.
Their main conclusion is that if IM-AGB stars were the
main polluters the current mass of the clusters should
be dominated by stellar remnants. Since this
does not appear to be the case, they consider
the winds of massive stars as a more attractive hypothesis.
Following up on this idea, \citet{decressin} investigated
the possibility that winds of {\em rotating} massive stars
are the polluters causing the observed abundance variations.
We shall later discuss these results   in the light
of our findings. 

\begin{table*}
\caption{Log of the observations}
\label{log}
\begin{tabular}{lcccccl}
\hline
\\
star \#& $\alpha$ & $\delta$  & date & UT & $t_{exp}$ & seeing \\
     & \multispan2{\hfill J2000\hfill}  & d/m/y  & h:m:s & s & arcsec \\
\\
\hline
\\
952 &00:21:39.14 & -72:02:53.73& 28/10/2001 & 00:37:18 & 6000 & 1\farcs{6}\\
952 & & & 28/10/2001 & 02:18:28 & 6000 & 1\farcs{3}\\
952 & & & 28/10/2001 & 06:14:43 & 2700 & 0\farcs{8}\\
975 &00:20:52.72 &-71:58:04.16 & 07/09/2000 & 06:28:57 & 5400 & 1\farcs{7}\\
975 & & & 07/09/2000 & 08:00:57 & 5400 & 1\farcs{7}\\
975 & & & 08/09/2000 & 05:42:13 & 3600 & 1\farcs{5}\\
975 & & & 08/09/2000 & 06:43:42 & 3600 & 2\farcs{4}\\
1012&00:21:26.27 & -72:00:38.73& 25/10/2001 & 03:31:42 & 6000 & 0\farcs{6}\\
1012& & & 25/10/2001 & 05:12:35 & 6000 & 0\farcs{8}\\
1081&00:21:03.82 &-72:06:57.74 & 29/08/2001 & 08:30:57 & 4500 & 0\farcs{7}\\
1081& & & 30/08/2001 & 07:26:16 & 4500 & 0\farcs{5}\\
1081& & & 30/08/2001 & 08:42:39 & 3600 & 0\farcs{7}\\

\\
\hline
\\
\end{tabular}
\\
\end{table*}

In this paper we 
explore the Li content of 47 Tuc, at 
the metal rich end of the metallicity range span by
halo GCs. In spite of its relatively high metallicity
47 Tuc is  very old (11.2 Gyr, \citealt{gra03})
and the Galactic  production should not
have greatly enhanced  its lithium content.

\begin{figure}[ht]
\resizebox{\hsize}{!}{\includegraphics[clip=true]{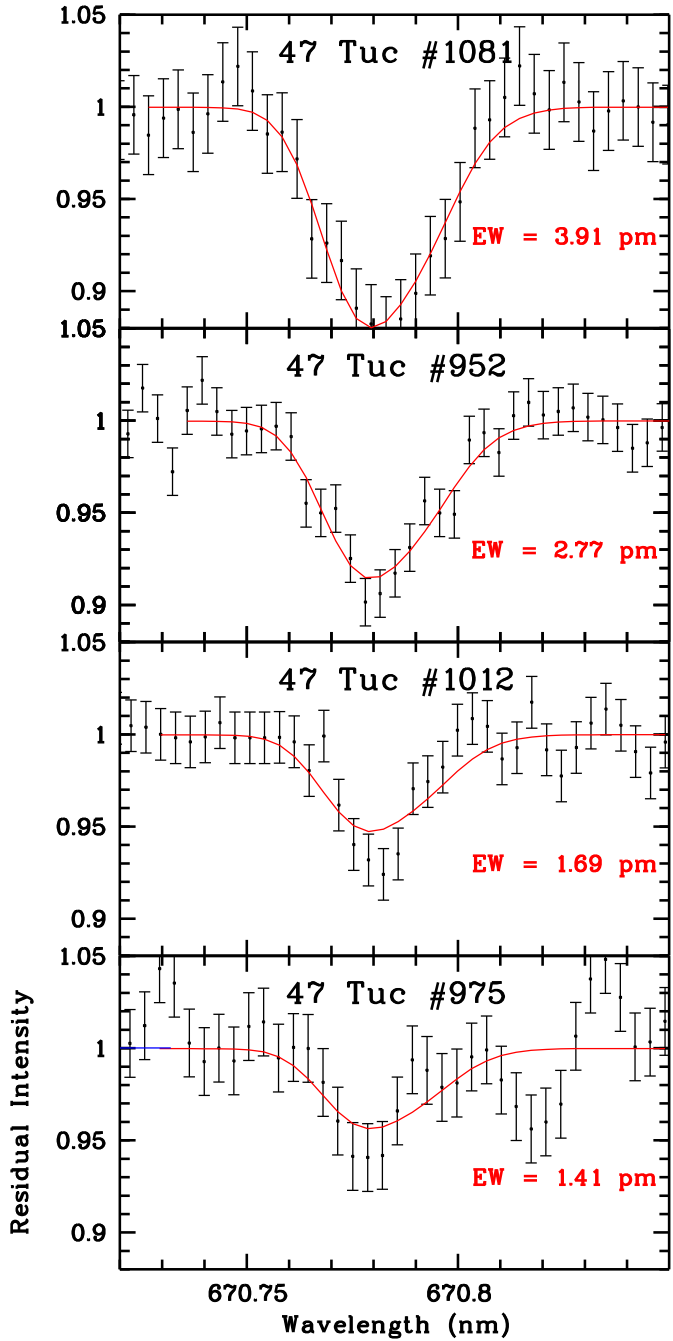}}
\caption{
\label{spectra}
Li doublet for the program stars, the best fitting synthetic profile
is shown as a thin line. }
\end{figure}

\begin{figure}[bt]
\resizebox{\hsize}{!}{\includegraphics[clip=true]{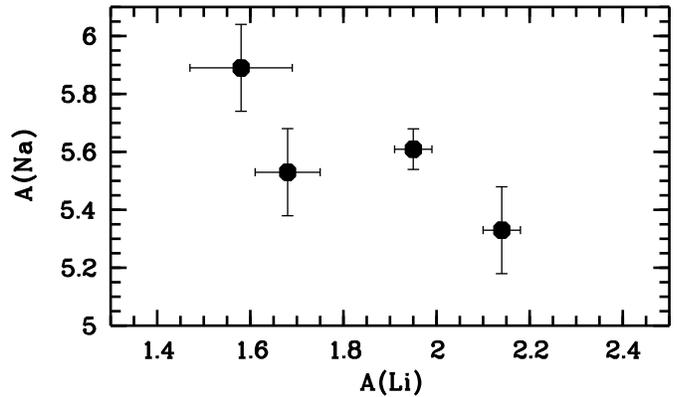}}
\caption{Na abundances versus Li abundances for the four stars 
measured by us.}
\label{lina}
\end{figure}

\section{Observations and data reduction}

Our spectra were collected at ESO-Paranal with 
the UVES spectrograph  \citep{Dekker}
at the Kueyen 8.2m telescope in the course of three runs,
covering two years. 
The log
of the observations is given in Table \ref{log},
the DIMM seeing was noted.
The data were reduced using the UVES context within MIDAS.
Different spectra of the same star  were co-added 
reaching S/N ratios in the range of 50 to 70 per pixel.
The coadded  spectra of the Li doublet for each star, together with the best
fitting synthetic profile are shown in Fig.\ref{spectra}.

\section{Lithium abundances}

The equivalent widths of the Li doublet were measured by
fitting synthetic spectra, as done by 
\citet{B02} and errors estimated through Monte Carlo simulations.
The chemical composition and atmospheric
parameters of these stars have been studied by
\citet{Carretta}. Both with respect to colours
and Balmer line profiles the four stars studied here are
twins and share the same effective temperature and surface
gravity. We adopt here the parameters of
\citet{Carretta}, namely \teff = 5832 K, log g = 4.05 (c.g.s. units)
and a microturbulent velocity of 1.07 \kms ~ for all the stars.
The iron abundance measured for these stars by \citet{Carretta}
is [Fe/H]$=-0.64$.
We used the ATLAS code \citep{k93,k05} to compute a  model atmosphere,
using the Opacity Distribution  Function (ODF) of
\citet{castelli} with
[M/H]$=-0.5$, $\xi=1$ \kms and $\alpha$ elements enhanced by 0.4 dex.
The procedure for 
Li abundance  determination was the same as in
\citet{B02}: we  iteratively computed 
synthetic spectra using the SYNTHE code \citep{k93,k05}
until the equivalent width of the synthetic spectrum matched 
the measured equivalent width.
The abundances are given together with the equivalent widths
in Table ~\ref{ews}. 
 
For star \# 952, for which  abundances are not given
in \citet{Carretta}, we measured  the Na abundance 
from the 616.1\, nm, 818.3\, nm and 
819.4\, nm lines, and derived log(Na/H)+12=A(Na)=5.61,
taking into account the NLTE corrections of 
\citet{gratton99}.

The only two other TO stars of this
cluster for which Li measures exist are 
the stars BHB 5 and BHB 7 (where BHB stands for
\citealt{briley}, who provide coordinates and finding
charts for these stars) that have been 
observed by \citet{pm97} with EMMI on the ESO 3.5m NTT 
telescope at a resolution R$\sim 18000$. 
These two stars have colours (see Table \ref{ews})
which place them in the same position in the colour-magnitude
diagram as the stars observed with UVES. We therefore decided
to use the equivalent widths measured by \citet{pm97}
and the same model atmosphere used for the stars observed with
UVES to derive the abundances provided in Table \ref{ews}.

\begin{table}
\caption{
\label{ews}
Equivalent widths and Li abundances for TO stars in 47 Tuc.
Errors take into account only the uncertainty on equivalent
widths, the effects of uncertainties in \teff ~ are neglected.}
\begin{center}
\begin{tabular}{lllccccrr}
\hline
\\
star \# & V & B-V&  EW  &  $\sigma_{MC}$  &S/N  & A(Li) & $\sigma_{Li}$ \\
        & mag& mag & pm  & pm              &     &       &  \\
(1)     & (2) & (3) & (4)       & (5) & (6) & (7) & (8)   \\
\\
\hline
\\
\multispan{8}{\hfill  measures from our  VLT-UVES data \hfill}\\
\\
\hline
\\
952 & 17.36 & 0.557 & 2.77   &  0.20 & 78  & 1.95 & 0.04 \\
975 & 17.33 & 0.597 & 1.41   &  0.30 & 54  & 1.58 & 0.11 \\
1012& 17.36 & 0.581 & 1.69  &  0.21 & 71  & 1.68 & 0.07 \\
1081& 17.37 & 0.587 & 3.91  &  0.24 & 47  & 2.14 & 0.04 \\
\\
\hline
\\
\\
\multispan{6}{\hfill measures of \citet{pm97}\hfill}\\
\multispan{6}{\hfill from NTT-EMMI data \hfill}\\
\\
\hline
\\
BHB 7 & 17.38 & 0.57& 5.30    &  0.80  &       & 2.30 & 0.08\\
BHB 5 & 17.35 & 0.59& 5.60    &  1.10  &       & 2.33 & 0.12\\
\hline
\end{tabular}
\end{center}
\end{table}

\section{Discussion}
 
\begin{figure}
\resizebox{\hsize}{!}{\includegraphics[clip=true]{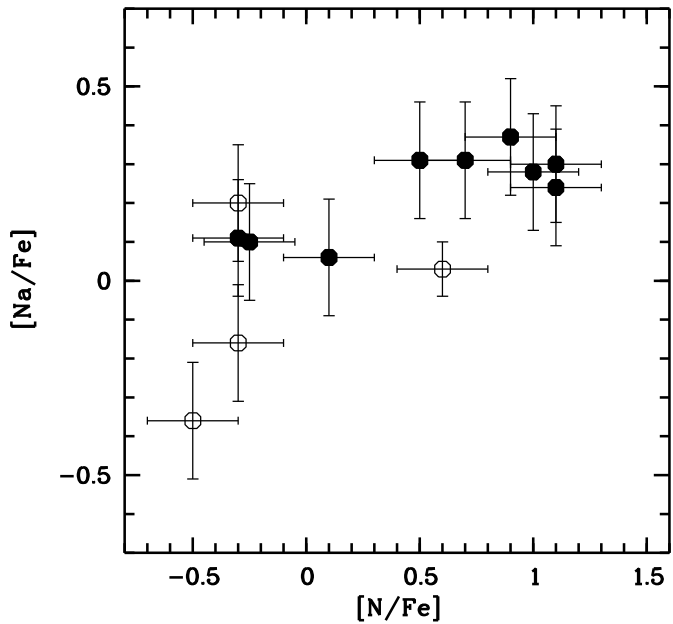}}
\caption{Na abundances versus N abundances for the four dwarf
stars (opens symbols) and for the
subgiant stars \citep{Carretta,Carretta05}.}
\label{nan}
\end{figure}

\begin{figure}
\resizebox{\hsize}{!}{\includegraphics[clip=true]{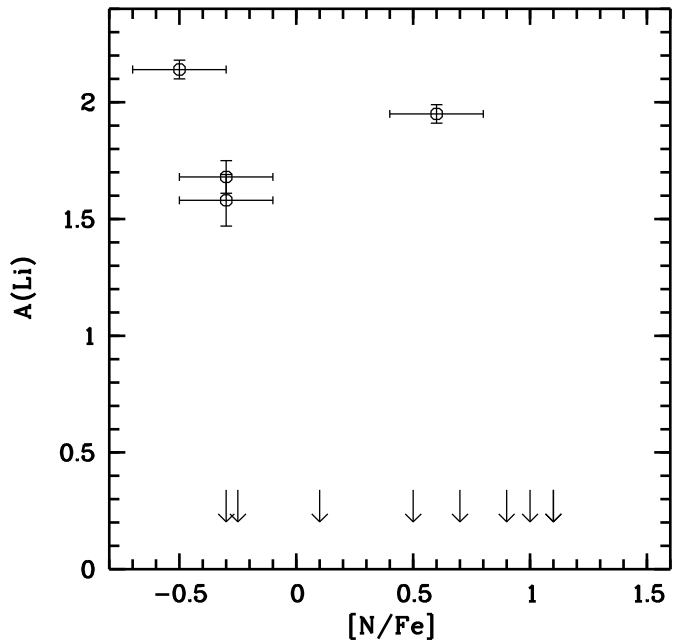}}
\caption{Li abundances versus N abundances for the four dwarf stars 
and for the subgiant stars \citep{Carretta,Carretta05}.
Only upper limits are available for the subgiants stars.}
\label{lin}
\end{figure}

\begin{figure}[t]
\resizebox{\hsize}{!}{\includegraphics[clip=true]{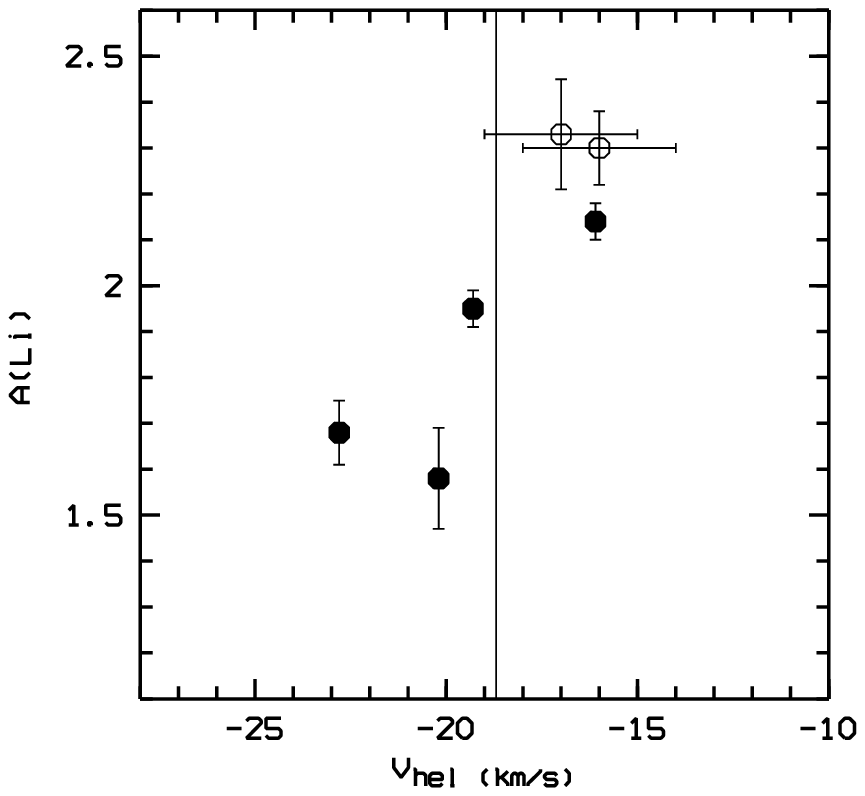}}
\caption{
\label{li_vhel}
Li abundances 
as a function of  heliocentric
radial velocities 
The four stars observed by us are shown as filled circles,
while the two stars observed by \citet{pm97}
are shown as open circles.
}
\end{figure}

\subsection{Are the Li variations real ?}

Although still very limited,
the data suggest that there is a real
scatter in the lithium abundances of this metal-rich
cluster. The mean Li abundance is 1.84 with a 
standard deviation of 0.25 dex.
We do not perform any correction for NLTE effects
or standard depletion as done by  \citet{B02} and \citet{BM92},
since all the stars have the same \teff ~ and these
would be the same for all the stars
and would have no effect on the dispersion in Li abundances.
A Monte Carlo simulation of 1000
ensembles of 4 "observations" with the errors reported
in Table \ref{ews}, as done by \citet{B02} and
\citet{BM92}, provides a mean dispersion
of 0.063 dex with a standard deviation of 0.032 dex,
we can claim that dispersion in excess of the
expected measurement error is detected at the 6 $\sigma$
level.  Since the errors in Table \ref{ews} arise
only from errors in the equivalent widths one could argue
that a real difference in the temperature of the stars
could justify the scatter observed.
A random scatter of 100 K in the effective temperatures 
would imply changes of the order of 0.08 dex in the 
derived Li abundances. If we increase all the errors
in Li abundances by this amount and perform another
Monte Carlo simulation the mean dispersion
is 0.141 dex with a standard deviation of 0.063 dex.
At this point the detection of extra scatter is marginal
at only 1.8 $\sigma$, yet still present. 
In order for the observed dispersion to be entirely
consistent with the observational errors, we would
have to increase the scatter in \teff ~up to 250 K.
Given the similarity of the spectra of the different
stars there is little support for the existence of such
a spread in the effective temperatures of our stars.
Furthermore, such a scatter in temperatures should
also produce some scatter in derived iron abundances,
which is not observed \citep{Carretta}.
We believe it is simplest to accept that the stars
observed by us do show real variations in the Li abundances.
Clearly the observation of a larger sample of stars
is required to firmly establish the reality
of Li variations.
We have repeatedly tried in the last four years 
to obtain time at the VLT
to observe further TO stars in this cluster,
but have been unsuccessful.

Let us assume, for the sake of discussion,
that the spread in Li abundances is entirely due
to observational errors. In this case it is legitimate to average
the values in Table \ref{ews} to obtain the Li content
in 47 Tuc. The mean A(Li) = 1.84
is 0.5 dex {\em lower} than the lithium content in 
NGC 6397. One is thus lead  to the inescapable conclusion
that Li has been depleted in this cluster.  
Under these assumptions the depletion would have been fairly uniform.

We believe it is less contrived to assume that Li has been depleted
in a non-uniform way and that this non-uniformity gives rise
to the scatter in Li abundances.
Finally one should not discard the information
provided by the two stars observed by \citet{pm97},
both of which show a higher A(Li) than any of the stars
observed with UVES, and indeed to a level
comparable to that observed in NGC 6397.
This considerably strengthens the claim
that there are indeed real variations in Li abundance
among TO stars in 47 Tuc.

\subsection{The Li-Na anti-correlation}

The
Li  abundances in 47 Tuc 
appear to be anti-correlated
with the Na abundances; in our view this is
a fact which further  supports
the reality of the abundance variations.
A plot of A(Li) versus
A(Na) is shown in Fig. \ref{lina}.
Kendall's $\tau$ test provide a probability of 82\%
that this anti-correlation is real. It is a bit
low to make a strong claim, but with only 4 points 
it is difficult to expect more definite results.
If the spread in Li abundances were due to incorrect
temperatures it could not possibly create such an
anti-correlation, since any change in \teff ~ produces
changes of {\em equal sign} and {\em comparable
magnitude} on both Li and Na.
There is no way to create the anti-correlation 
by an incorrect choice of atmospheric parameters.

If we accept the hypothesis that the ``unpolluted''
Na abundance of the cluster is provided by the stars
with the highest Li abundance then
it  should
be [Na/Fe]=--0.36, or perhaps even
lower, if the stars BHB 5 and BHB 7 follow
the trend traced by the other four stars. 
We examined again the EMMI spectra of
the two stars observed by \citet{pm97}
to see if it were possible to measure 
the Na abundance from those spectra.
The only usable Na lines 
were those of the 615.4--616.0\,nm doublet,
which are weak in these warm TO stars.
Given the low S/N of the spectra, the lines cannot be reliably used and only 
a   very high (not significant) upper limit can be obtained.

\subsection{The need for  ``pollution''}

When considering field stars, the usual interpretation 
of relatively cool and metal-rich stars found below
the {\em Spite plateau} is that Li is  depleted in these
stars due to a deeper convection zone and/or  atomic diffusion
\citep{michaud84,vc95,vc98,salaris,ric3,ric1,richard}.
In the case of 47 Tuc the observation of low Li abundances
is accompanied by the finding of the Li-Na anti-correlation,
which implies that  some of the polluting 
material has  been processed
at temperatures in excess of $3\times 10^{7}$ K so that 
Li has been destroyed and Na created by $p$ captures on
$^{22}\rm Ne$. These temperatures, are high enough 
for extensive burning of oxygen through the CNO cycle, 
which can well explain
the Na-O anti-correlation present in this
cluster \citep{Carretta}.
These temperatures are however too high to be found
within TO cluster stars, which, even at the centre
should not exceed a temperature of $2\times 10^7$ K.
Therefore the Na-O anti-correlation requires processing
in a previous generation of stars
and subsequent non-uniform pollution of the ISM.
This view has to be consistent 
with  the [N/O] ratio in the TO stars
in this cluster, which has a mean value of -0.85 dex
\citep{Carretta05}, that  is about 0.5 dex {\em lower}
than what observed in field stars
of the same metallicity \citep{israelian04}.
Also the $^{12}$C/$^{13}$C ratios ($> 10$, \citealt{Carretta05})
argues against CNO cycling of material in the TO stars
of 47 Tuc.
This poses serious problems to explain the
observed Li-Na anti-correlation.
The  polluting material probed by the abundances
in the TO stars of 47 Tuc, must have 
experienced  high temperatures in order  
to produce the Li-Na anti-correlation, 
however we do not observe signature of
extensive O burning, N production
or $^{13}$C production.
A further complication in this
picture is the possible {\em production}
of Li, e.g. via Cameron-Fowler mechanism
\citep{CW} or otherwise.
Any production of Li, however, would tend to erase
the Li-Na anti-correlation, it is therefore likely
that there is no, or very little, Li production.

In Fig. \ref{nan} we show [Na/Fe]
as a function of [N/Fe] for both the dwarf
and subgiant stars observed in 47 Tuc.
There is a hint of a correlation, albeit with
a very large scatter.
We inspected all the spectra of
the subgiant stars observed in this cluster with UVES, but
could not convincingly detect the Li doublet in any of them.
Considering the quality of the spectra we set an upper
limit on the equivalent width of the Li doublet in these
stars of 1 pm, this implies an upper limit
A(Li)$< 0.34$.
In Fig. \ref{lin} we show the Li abundances, or upper limits,
as a function of [N/Fe], contrary to what observed
in the cluster NGC 6752 \citep{Pasquini6752}, 
Li and N appear to be totally uncorrelated.
We stress that some caution must be exerted
in interpreting this data since too few stars
have been observed. 
It should be noted that according to the measures of \citet{briley},
BHB 5 is CN-weak CH-strong, while BHB 7 is CN-strong  CH-weak
(see Fig. 7 of \citealt{briley}).
If we interpret the CN band strength in terms of N abundance,
this would be further evidence of what is hinted
at by Fig. \ref{lin}: that Li abundance does not seem to 
be correlated to N abundance.
It would  clearly be of great interest to re-observe
stars BHB 5 and BHB 7 with an 8m class telescope
in order to perform a complete chemical analysis and
improve their Li abundances.

\subsection{Neutron capture elements: absence of an AGB signature}

It is interesting to look also at the abundances
of the $n-$capture elements in these stars,
which have been measured by \citet{gael}.
In this respect the cluster seems extremely 
homogeneous and the abundances are 
consistent with those observed in field
stars of the same metallicity, with the exception
of Sr, for which both the [Sr/Fe] and the
[Sr/Ba] ratios are slightly {\em higher}
than in field stars of the same metallicity.
This situation makes it unlikely that these
stars have been formed out of, or polluted by,
material heavily enriched by $s-$ process
elements, as may be expected in the ejecta
of AGB stars which have undergone thermal pulses.

It should be here noted that  a 
recent investigation of the chemical
composition of AGB stars in this
cluster \citep{wylie} has claimed a true
scatter in the ratios of $n-$capture elements.
Since the stars currently observed on the AGB
in 47 Tuc are of too low mass to have undergone
the third dredge-up one should conclude that
this inhomogeneity is intrinsic and not
due to the self--pollution. To further support
this point of view \citet{wylie} point out  
the large scatter in sodium abundances in
their stars ($+0.3 \la $[Na/Fe] $\la +1.0$)
and claim  that this, and the variations in
$n-$capture elements could be explained by
the presence of at least two separate
stellar populations.
This suggestion is intriguing, however
the results of \citet{wylie} for $n-$capture 
elements are at odds with those of \citet{gael}
for the stars examined in the present paper
and with those of \citet{alan} for a sample
of 5 giants stars in this cluster.
Further investigation would be desirable
to rule out the possibility of systematic
differences among the different analysis.
Note that the Na enhancements
observed by \citet{wylie} are all  larger than
what observed in our TO stars.

\subsection{Do all GCs evolve in a similar way ?}

It appears clear that GCs exhibit a considerable
diversity in their abundance inhomogeneities.
NGC 6397 displays no Li--Na anticorreation and
a marked enhancement of N.
NGC 6752 displays a well defined Li--Na anti-correlation,
accompanied by a rather large enhancement
of nitrogen, compared to field stars,
and a [N/O] ratio which  ranges between 0.6 and 1.8, over
two orders of magnitude larger than what observed in
field stars of the same metallicity.
47 Tuc displays a Li--Na anti-correlation, however 
N does not appear to be enhanced and subsolar values
of [N/O] are found.
Recently \citet{bekki} have proposed a scenario by which
GCs are formed at high redshift
in dwarf galaxies, embedded in dark matter subhalos and
the polluters are mainly the field IM-AGB stars of the host galaxy,
which is subsequently tidally disrupted.
Such a scenario has several appealing aspects, among which,
in our view, the most interesting is that it may accomodate
quite naturally differences in self-enrichment histories
of GCs, which may be traced to the
different properties (masses, dynamics, age...) of the,
now disrupted, host galaxies and dark matter subhalos.
Currently such models are unable to explain the Na-O
anti-correlations, and, by inference, they should
likewise be unable to explain the Li-Na anti-correlation.
However, more generally, the abundance pattern in
the GC hosting dwarf galaxies may result
from complex histories, which may allow to explain
the variety of abundance patterns observed in GCs.
The scenario of \citet{bekki} is supported by
the dynamical simulations of \citet{gnedin}, which
suggest that GCs may form in giant molecular
clouds within high redshift galaxies. By computing
the orbits of such clusters in a Milky Way-sized galaxy
\citet{gnedin} conclude that all clusters found 
at distances larger than 10 kpc
from the Galactic center were indeed formed in 
satellite galaxies,
which have now been tidally disrupted. 

At this point it is perhaps worth to mention 
the significative difference in HB morphology
between NGC 6752 and 47 Tuc, the former beeing
characterized by an extended blue tail.
Of course this difference could be simply due
to the different metallicity of the two
clusters, however
the long blue tail in the
HB of NGC 6752 could also be 
linked to stars polluted by He-enriched matter,
according to the scenario suggested by \citet{DAC}.

\subsection{Are massive stars the polluters ?}

The lack of nitrogen enhancement seems difficult
to reproduce using AGB stars as polluters.
A distinct possibility is that the
polluters are instead massive stars, an
alternative which is favoured by  \citet{prantzos}.
These authors suggest that it is the
wind of massive stars which is retained in the cluster,
while the SN ejecta are lost, due to their 
higher speed of ejection. 
\citet{meynet}  present the results for
models of 60 M\sun\ star of very low
metallicity ($Z/Z_\odot = 10^{-8}$ and
$Z/Z_\odot = 10^{-5}$), both with
and without rotation.
According to these computations the {\em wind} of 
such a star, can provide large $^{12}$C$/^{13}$C ratios,
however very small N/O ratios, as observed in this cluster,
only if the
star has a low enough rotational velocity, so that
rotational mixing is unimportant.
If instead also the SN ejecta are mixed to the wind,
due to the large production of O, the N/O ratio
is considerably lowered, whatever the rotational
velocity. It seems however likely that the fast
S/N ejecta are lost to the cluster and do not
contribute to its chemical evolution, at variance to what
we expect from the relatively slow-moving winds.
The investigation of \citet{meynet}
has been extended by \citet{decressin}, who 
computed also models for 20, 40, 60 and 120 M\sun,
for a metallicity of $Z=0.0005$, which 
corresponds, roughly, to [Fe/H]=--1.5, adequate, e.g.
for NGC 6752. These rotating models seem to produce
winds with a composition apt to reproduce
the C,N, O and Na variations in NGC 6752.
We cannot apply directly these models to 47 Tuc,
which is considerably more metal-rich. We note 
however that the winds of mass 60 and 120 M\sun, 
according to \citet{decressin}, display low 
N/O ratios only up to the end of central H-burning,
after this phase N/O is always greater than the
solar ratio, which is $\sim 0.1$, while the observed  N/O
ratio in 47 Tuc is lower than solar.
The 20 M\sun\ model never provides wind with 
N/O $>1$ and the 40 M\sun\  does so only after
the appearance of the He-burning products 
at the surface. \citet{decressin}
do not provide the fraction of  $^{13}$C
in the wind, so that we cannot tell if
at any of these phases a low N/O is
accompanied by high $^{12}$C/$^{13}$C.
Even if it were so, one would have to admit
that the pollution was made only
during such phases (or by stars of such masses, e.g.
masses of 20 M\sun\ or lower).
\citet{decressin}
assume that the massive stars winds are essentially
Li free and invoke a dilution of the wind with
about 30\% of pristine gas, in order to reproduce
the lowest Li abundance observed in NGC 6752,
assuming the pristine lithium was 
what derived from the SBBN predictions, 
and the baryon to photon ratio provided by WMAP.
In the case of 47 Tuc, since some of the observed values
of Li are even lower than in NGC 6752, these
should have been formed almost exclusively
out of  the winds, with an addition of at most 9\%
of pristine material.
It is certainly true that at the 
temperature necessary  for sodium production
the lithium should be completely destroyed,
therefore to be consistent with the observed
Li abundances, the processed material must
anyway be diluted with material in which Li has been 
preserved. 
Such a pollution may provide the observed
Li-Na anti-correlation, however the abundances of
N and $^{13}$C should follow this pattern.

\subsection{A kinematical signature of pollution ?}

An interesting feature emerges if we plot Li abundances
as a function of radial velocities (see Fig.\ref{li_vhel}): there is a 
mild hint (probability of correlation between
radial velocity and Li abundance
$\sim 91$ \%)
that the most Li-rich stars have a radial velocity different
from the  less Li-rich.
This may point towards a kinematic distinction
between the more polluted and the less polluted
stars, however,
given the limited size of the sample  it is premature 
to claim this is a real feature; nevertheless it surely prompts
for the observation of a larger sample of stars.

\balance
\section{Conclusions}

The new high quality spectra of the Li doublet in
TO stars of 47 Tuc strongly suggest that Li has
been depleted in a non-homogeneous way in this cluster.
The existence of an  anti-correlation
between A(Li) and A(Na) effectively rules out the
possibility that the spread in A(Li) arises from
incorrectly chosen atmospheric  parameters.

The newly established 
Li-Na anti-correlation, and the
Na-O anti-correlation argue in favour of 
nuclear processing which has taken place in a previous generation
of stars.
The low nitrogen abundance in these stars places
however a strong constraint on the properties
of this previous generation of stars.
It appears unavoidable to conclude that the
self-enrichment history of 47 Tuc has been distinctly
different from that of the more metal-poor cluster NGC 6752.
\begin{acknowledgements}
We are grateful to L. Piau for interesting discussions
on the topic of Li depletion.
PB and LS acknowledge support from the MIUR/PRIN 2004025729\_002 and from EU
contract MEXT-CT-2004-014265 (CIFIST).
\end{acknowledgements}

\bibliographystyle{aa}

\end{document}